# Recent advance in phase transition of vanadium oxide based solar reflectors and the fabrication progress

Golsa Mirbagheri[1], David T. Crouse[2], Chee-Keong Tan[2]

*Duke University Pratt School of Engineering[1]*
*Computer and Electrical Department of Clarkson University[2]*

**ABSTARCT**

Vanadium dioxide ($VO_2$) as a phase-change material controls the transferred heat during phase transition process between metal and insulator states. At temperature above 68 ̊C, the rutile structure $VO_2$ keeps the heat out and increases the IR radiation reflectivity, while at the lower temperature the monoclinic structure $VO_2$ acts as the transparent material and increase the transmission radiation. In this paper, we first present the metal-insulator phase transition (MIT) of the $VO_2$ in high and low temperatures. Then we simulate the meta-surface $VO_2$ of metamaterial reflector by Ansys HFSS to show the emittance tunability ($\Delta\varepsilon$) of the rutile and monoclinic phase of the $VO_2$. In next section, we will review the recent progress in the deposition of thermochromic $VO_2$ on glass and silicon substrate with modifying the pressure of sputtering gases and temperature of the substrate. Finally, we present the results of the in-situ sputtered $VO_x$ thin film on thick $SiO_2$ substrate in different combination of oxygen and argon environment by $V_2O_5$ target at temperature higher than 300 ̊C and then, analyze it with x-ray diffraction (XRD) method. The thermochromic $VO_2$ based metamaterial structures open a new route to the passive energy-efficient optical solar reflector in the past few years.

## 1. INTRODUCTION

The reversible metal-insulator phase transition (MIT) of $VO_2$ has gained enormous attention in thermal control system of spacecrafts recently. However, among other metamaterials great effort is still investigated to understand the MIT process. In this section we review the recent advance in MIT mechanism of $VO_2$ and discuss the physical properties of the rutile and monoclinic structures. V-O system is one of the most interesting materials with different compounds including $VO$, $VO_2$, $V_2O_3$, $V_3O_5$, etc. Some of these components show metal-insulator transition (MIT) with a big change in thermal, optical, or electrical properties. These special properties encouraged scientists to use the V-O system in many optical applications and therefore, learn the phase transition process. Among all various types of V-O system, $VO_2$ is considered for the temperature phase change mechanism [1, 2, 3]. Today, great effort has been made to understand the MIT phase change mechanism on various $VO_2$ morphologies acquired from different types of fabrication methods. [4, 5, 6]. Despite the great progress has been made in recent years, the phase transition is still an arguable mechanism. The rapid transition between rutile metallic phase and monoclinic insulator phase is followed by changes in the crystallographic and electrical properties of $VO_2$, accomplished by switching and sensing applications [7].

The isolator behavior of $VO_2$ with $3d_1$ orbital occupancy in room temperature and, forming other isolator phases in specific temperatures, all make it hard to interpret the phase change properties. One explanation for this change can be described by Peierls mechanism and electron correlation technique, which address the lattice distortion and the band gap shift. On the other hand, in the 3d t2g-derived states of the vanadium, when $d_l$ bands break into occupied bonding and empty antibonding, the π* states get depopulated and their energy increase that leads to create the bandgap. Therefore, a bandgap is created between the d$_∥$ and the π* bands, illustrated in Figure 1. However, bandgap disappear in rutile phase to allow reflection of radiations. Therefore, we explain the lattice structures and electron interaction to learn more about the transition process.

---

[1] Send correspondence to Golsa Mirbagheri
E-mail: gm226@duke.edu

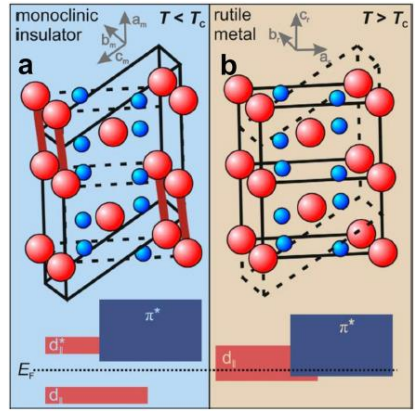

Figure 1. Bandgap in monoclinic (left) and rutile (right) phase of V *[1]*

The $VO_2$ can grow in the form of different crystal structures, including monoclinic (M) and rutile (R) phases which show reversible phase change close to room temperature. During the phase change, the d-electrons of V atoms in the symmetrical structure join to the varying V-V bonds in low symmetry monoclinic structure. There is an assumption that this V-V localized structure derived from the high temperature delocalized phase and shows insulative properties. On the other hand, the conductivity property in the high temperature rutile phase is due to the fermi level between the π* band and the d$_∥$ band. While in M phase, the d$_∥$ band split to two parts, make a gap between d$_∥$ band and the π* band where Fermi level falls into that and lead to $VO_2$ to be insulative. However, this theory is no clarified yet, since some research show that lattice distortion is not the only reason for MIT phase transition. One hypothesis is that there are other phases rather than M, at the low temperature. With the help of doping, electron injection or stress effect, different phases of $VO_2$ like M2 monoclinic, triclinic $VO_2$ (T) or other metastable monoclinic metal phases are discovered, which indicated that V-V bond interactions is not the necessary step to open an insulating gap [1, 8, 9, 10]. Studying the transition phase of thermochromic $VO_2$ makes a new path to understand the correlation effects of other thermal sensors [11, 12, 13].

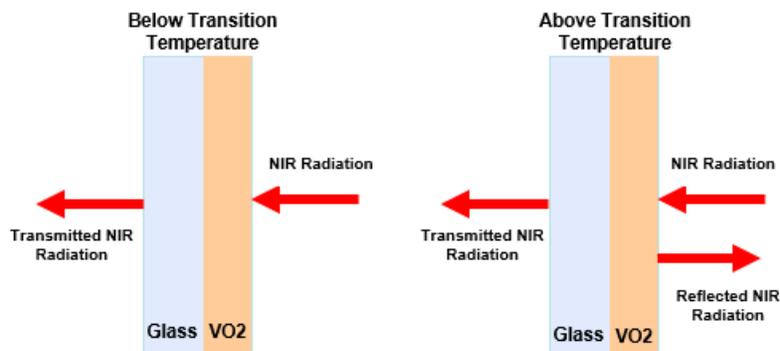

Figure 2. Vanadium dioxide semiconductor (left) to metal (right) phase transition at 68 ℃ shows change in infrared reflectivity

As a thermochromic material, $VO_2$ goes through semiconductor to metal phase transition at the specific temperature of 68 ℃ which makes the change in infrared reflectivity, as shown in Figure 2. The transmission radiation changes with temperature and thickness of the $VO_2$ material, which is demonstrated in Figure 3. The broadest extensive hysteresis transition belongs to the thickest $VO_2$ with 50nm. By increasing the temperature above 68 ℃ at wavelength of 2 µm, the material keeps the heat out and IR radiation reflection increased since little radiation is transmitted into. To manipulate the transition temperature of the $VO_2$ film different method were introduces including the adding various dopants, using multiple layers, or size effectiveness of $VO_2$ lattice [14, 15, 16, 17].

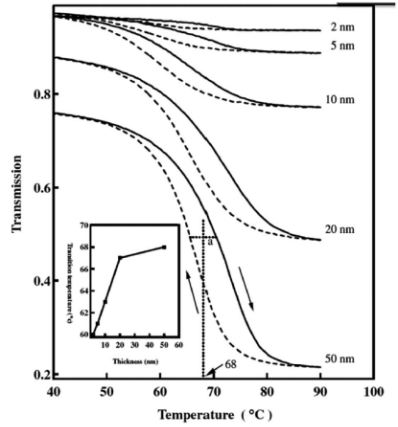

Figure 3. Transmission hysteresis loops for different thickness of $VO_2$ thin film at wavelength 2µm *[14]*

## 2. NUMERICAL SIMULATION OF FILTER

In [18], the optical solar reflector (OSR) composed of plasmonic $VO_2$ meta-surface is introduced to control the temperature of the satellite devices. The grating thermochromic $VO_2$ in the OSR filter improve the absorption (α) and emittance tunability Δε with the help of plasmonic and thermochromic properties of $VO_2$. This leads to higher absorption of the structure in higher temperature and lower absorption in temperature below the critical level. The smart meta-surface reflector shows higher tunability Δε at lower absorptance in comparison with the planar thin film design [19]. The OSR reflects the radiation of the sun and dissipates the thermal spectrum of onboard instruments to reduce the solar absorption and keep down the thermal fluctuation of the satellite. In contrast with active systems, the smart passive systems keep the heat at low temperature (low emittance) and dissipates extra heat at high temperature (high emittance) to keep the temperature at optimal ranges. This makes the passive systems cost and energy effective with lower weight. In passive thermal control structures, thermochromic $VO_2$ works as a reflector and shows semiconductor to metal phase change at specific temperature 68C with high tunability of emittance. The critical temperature can be manipulated by doping or defect engineering methods to be near room temperature. The performance of the passive smart system can be explained by the dynamic emittance tunability of $\Delta\varepsilon = \varepsilon_{hot} - \varepsilon_{cold}$. $VO_2$ based metamaterials exhibits phase transmission on plasmonic effects that shows higher absorption in metallic state and lower emittance in monoclinic phase, applicable over a wide range from visible and NIR to MWIR.

The insulator $SiO_2$ in [18] is sandwiched between the meta-surface $VO_2$ and aluminum reflector, shown in Figure 4, makes the destructive interference effect to improve the blackbody spectrum absorption and high tunability Δε. The $Al_2O_3$ layer is considered as adhesion material.

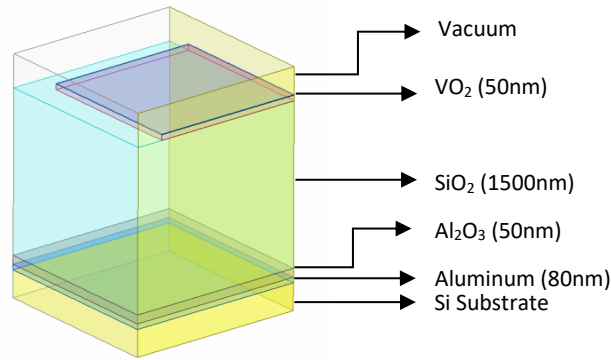

Figure 4. $VO_2$ based OSR Schematic Simulated in HFSS

The proposed structure is composed of the array of metasurface $VO_2$ in the shape of the square grating and simulated by Ansys HFSS, demonstrated in Figure 4. The absorption infrared spectra of the monoclinic grating $VO_2$ with grating squares size 2.8μm and different gaps between grating square changes from 3.3-4.35μm at temperature 30°C is low at 5-7μm and 12-18μm which is illustrated in Figure 5.

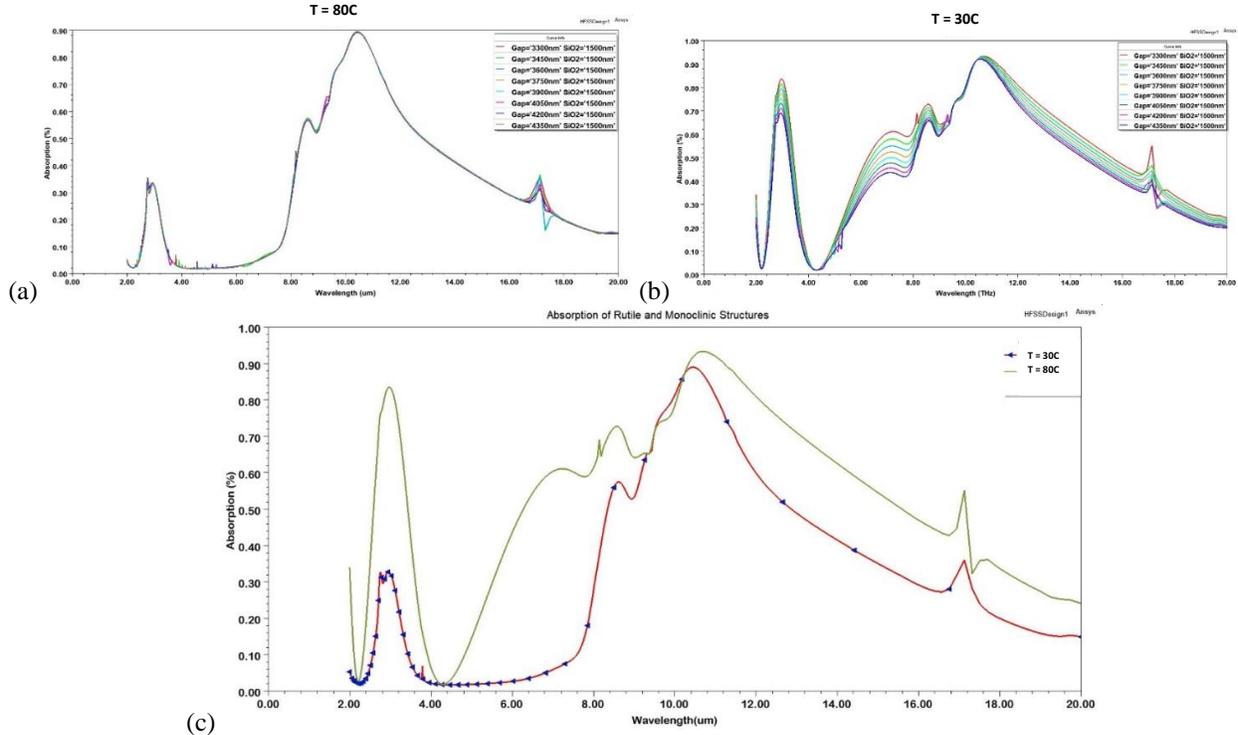

Figure 5. Absorption of meta surface $VO_2$ at 30°C (a) and at 80°C (b). Comparison of absorption of meta surface $VO_2$ between two rutile and monoclinic structures of grating $VO_2$ based metamaterial structure (c)

The emittance of a reflector is described in Eq. 1 as the average emittance weighted by blackbody spectrum at temperature (T).

$$\varepsilon = \frac{\int(1-R(\lambda))B(\lambda,\ T)d\lambda}{B(\lambda,\ T)d\lambda} \quad (1)\ [18]$$

Using Eq 1, the emittance of the meta-surface $VO_2$ with low-emissivity dielectric spacer in high and low temperature is calculated by the measuring the reflection spectra from FTIR method, as shown in Figure 6 (a). Therefore, it's not hard to find the emittance tunability Δε of the meta-surface $VO_2$ at the peak of the blackbody spectrum, which is plotted in Figure 6 (b). The latter shows that by increasing the gap size between features (or decreasing the feature size of the grating $VO_2$), maximum emittance tunability Δε improved about %30 in comparison with the planar film of $VO_2$.

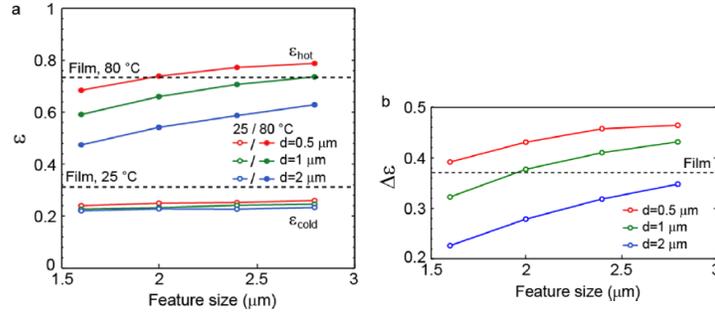

Figure 6. Measured emittance (left) and emittance tunability (right) of planar film (dash line) and meta-surface VO$_2$ for high and low temperatures [18]

## 3. REVIEW OF RECENT VO$_2$ DEPOSITION WORKS

In the recent years, different methods were advised to sputter vanadium films [20, 21, 22]. In order to make thermochromic VO$_2$ of metamaterial structures, we review some deposition approaches with the help of modifying the pressure of gases and temperature of the substrate [23, 24, 25]. One method is to sputter thin film with vanadium target and then oxidized it in high temperature annealing environment at temperature 300 ℃ or above. The other popular way is reactive sputtering deposition which is performed with the partial oxygen pressure %0 to %20 of the total pressure (argon/oxygen atmosphere) at 300-600 ℃ to create different mixtures of VO$_x$ (V$_2$O$_3$, V$_2$O$_5$ or VO$_2$) on the substrate. The films can be analyzed with XRD, atomic force microscopy and X-ray photon spectroscopy (XPS) methods, in addition to checking the color of different composition of VO$_x$ visually [26]. Furthermore, post annealing in 300 ℃ or higher is recommended to form various types of VO$_x$ in other works. In [23], different forms of VO$_x$ were deposited on substrate with RF magnetron deposition by V$_2$O$_5$ target at temperature 300 to 527 ℃ without annealing process. At the oxygen partial pressure less than $10^{-33}$ in room temperature in RF magnet deposition process, the VO$_2$ is the most stable phase of the other composition of VO$_x$ from the V$_2$O$_5$ target. The VO$_2$ thin film deposited on the glass substrate at 125W plasma power with the oxygen partial pressure changing from 0-20% of argon/oxygen for 100 minutes. Meanwhile, at the oxygen partial pressure higher than $10^{-29}$, V$_2$O$_5$ grows on the substrate as shown in Figure 7 [23, 27, 28].

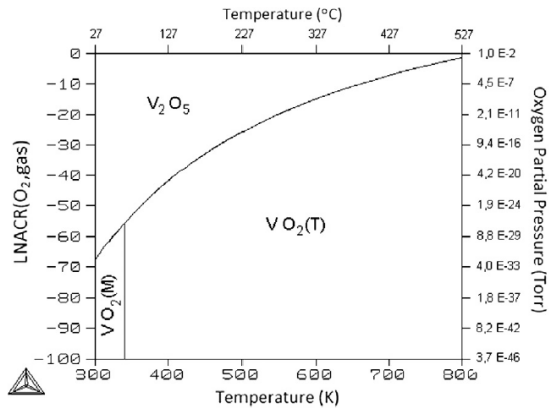

Figure 7. Monoclinic (M) and tetragonal (T) VO$_2$ and V$_2$O$_5$ thin films growth from V$_2$O$_5$ target at partial oxygen pressures *[23]*

Reactive hydrogen is used for Rf sputtering of vanadium film with V$_2$O$_5$ target to reduce oxygen, whereas reactive oxygen is suggested for metallic vanadium or V$_2$O$_3$ target. The V$_2$O$_3$ target provides broader flow ratio during sputtering vanadium oxide, but it is more expensive than other two targets. Furthermore, metallic target gets oxidized during sputtering process, therefore, V$_2$O$_5$ is the most stable and best sputtering target to deposit rutile VO$_2$. However, due to safety requirement, reactive oxygen is recommended with V$_2$O$_5$ target at higher temperature. In [24], rutile VO$_2$ is deposited on fused silica glasses, Si or thick SiO$_2$ on Si substrate from a V$_2$O$_5$ target by tuning substrate temperature and the oxygen flow rate. The temperature of substrate changed from 300 to 500 ℃ during the deposition with the total oxygen/argon pressure 12.5 mTorr at the power of 60 or 120W and the oxygen flow ratio ($R_{fo}$) is adjusted between 0.042 to 0.1 based on Eq 2, where f is the flow rate of the gas.

$$R_{fo} = \frac{f_{O2}}{(f_{Ar} + f_{O2})} \quad (2)\ [24]$$

The temperature of the substrate plays an important role for adjusting the oxygen component of the thin film. Figure 8 shows at lower temperature 300 °C, the $V_4O_9$ phase starts to grow on Si substrate. As temperature increases, $V_6O_{13}$ and meta-stable phase $VO_2$ (A) phases show at $R_{fo} = 0.1$ due to less oxygen content of the films. Other phases with lower oxygen grow on the Si substrate when $R_{fo}$ is decreased. At higher temperature 500°C, only rutile $VO_2$ grows on the Si substrate with $R_{fo}$ between 0.06 and 0.042 in power 60W.

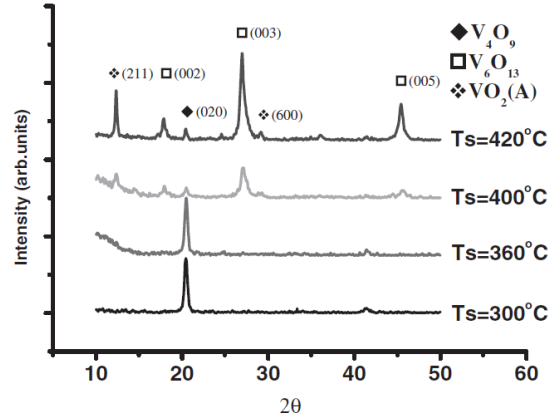

Figure 8. XRD result of thin films on Si substrate at different temperature from 300°C to 420°C for $R_{fo}$ = 0.3 [24]

Figure 9 shows that the increasing the temperature of substrate not only form the phases with lower oxygen, but also increase the average grain size of the deposited thin film.

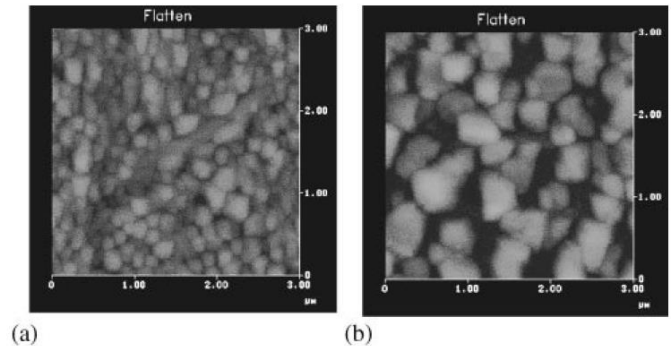

Figure 9. AFM results for grain sizes of mixed rutile $VO_2$ and $VO_2$(A) phases at T = 450°C (left) vs single phase rutile $VO_2$ film at T = 500°C (right) *[24]*

In [25], $VO_2$ is sputtered from $V_2O_5$ target with reactive hydrogen (2.5-10% of $H_2$+Ar) at pressure 0.2 Pa and power 100W. After 30 min pre-sputtering, the temperature of the fused silica glass substrate kept at 400 °C to grow the $VO_2$ smoothly. In [29], the deposited precursor vanadium on the $SiO_2$/Si substrate with metal vanadium target is oxidized to $VO_2$ in chamber at temperature 370-415°C and oxygen pressure 0.2Torr. Furthermore, post annealing at 270 °C (better result at 450 °C) is an alternative method to oxidize the deposited vanadium to shape the $VO_2$. However, the best results were related to the in-situ deposition immediately after deposition of precursor vanadium. In [30], the single (001) orientation orthorhombic metal vanadium is in situ deposited on sapphire, silicon, and glass substrates by RF sputtering with vanadium target in partial oxygen pressure 0-25% at temperature less than 100°C. In [31], the deposited vanadium post-annealed at 530°C with an oxygen pressure between 14 -18 Pa to not oxidize changing to

$V_2O_5$. In [32], the phase transition of $VO_2$ is improved by further post-annealing at 300°C in high-pressure oxygen 10–25 mTorr. The XRD result in Figure 10, shows thin film $VO_2$ sputtered by pulsed laser deposition (PLD) method on thick oxide ($SiO_2/Si$) substrate [33].

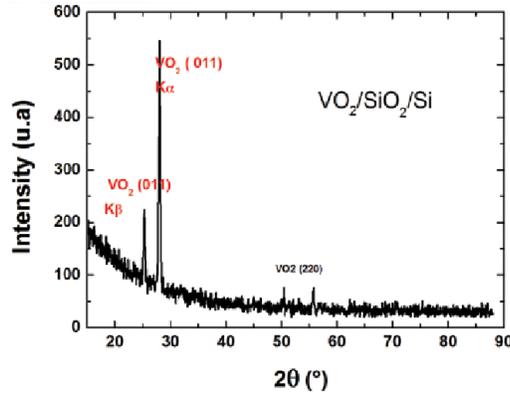

Figure 10. XRD results for $VO_2$ film deposited on $SiO_2/Si$ substrate *[33]*

## 4. EXPERIMENTAL RESULT

In our work, we sputtered 50nm $VO_x$ film on $SiO_2/Si$ substrate by $V_2O_5$ target. The substrate is composed of 1.5μm thermal $SiO_2$ deposited on silicon wafer by plasma-enhanced chemical vapor deposition (PECVD) method. During RF sputtering process, total gas pressure was kept at 12.5 mTorr for 2hours deposition process. We did several sputtering tests for different temperature of substrate varied from 300°C to 500°C in-situ at a power of 120W for different oxygen and argon flow ratio. The wafer height in the chamber held 20cm from target. The surface of the thin film is characterized by XRD at temperatures 300°C, 400°C and 500°C in Figure 11, shows that for flow ratio $O_2$= 2sccm and Ar(P) = 32sccm ($R_{fo}$ 0.058), the deposited film on thick $SiO_2$ is $V_2O_3$. We repeated the test for $R_{fo}$ between 0.042 to 0.06, however the measured surface of the thin film by XRD technique showed $V_2O_3$ growth on the thick oxide substrate. The result did not change significantly after post-annealing process for partial oxygen pressure 100Torr at 300°C, 400°C and 500°C. For the total volume of oxygen and nitrogen equal to 10 lit/min, %13 oxygen (1.3 lit/min) and %43 nitrogen (8.6 lit/min) were considered to do post-anneal the sputter film about 10 minutes. In [34], the ALD deposited amorphous vanadium oxide crystallized in furnace annealing process in higher rate of nitrogen ratio (more than %98) at temperature of 425°C and higher, as shown in Table 1 [35, 36, 37, 38]. The gray rows in Table 1 show the successful annealing process of create polycrystalline stoichiometric $VO_2$ phase.

| Anneal Temp | Air | Annealing time and gas ratio under atmospheric pressure | | | | Annealing time and gas ratio under low vacuum ($10^{-2}$ Torr) | | |
|---|---|---|---|---|---|---|---|---|
| | | $N_2$ (100%) | 98.78% $N_2$ + 1.22% $O_2$ | 99% $N_2$ + 1% $O_2$ | 98.5% $N_2$ + 1.5% $O_2$ | 98.2% $N_2$ + 1.8% $O_2$ | 99.4% $N_2$ + 0.6% $O_2$ | $2.7 \times 10^{-2}$ Torr (1 sccm $O_2$) |
| 420°C | 30 min | | | | | | | |
| 420°C | | 30 min | | | | | | |
| 420°C | | | | 5 min | | | | |
| 420°C | | | | | 5 min | | | |
| 420°C | | | | | | 10 min | | |
| 420°C | | | | | | | 10 min | |
| 425°C | | | 30 min | | | | | |
| 430°C | | | | 5 min | | | | |
| 450°C | 30 min | | | | | | | |
| 450°C | | 30 min | | | | | | |
| 450°C | | | 30 min | | | | | |
| 450°C | | | | | 5 min | | | |
| 450°C | | | | | | 10 min | | |
| 450°C | | | | | | | 10 min | |
| 500°C | | | | | | | | 60 min |

Table 1. The furnace annealing result to crystallize the ALD deposited amorphous vanadium oxide film *[34]*

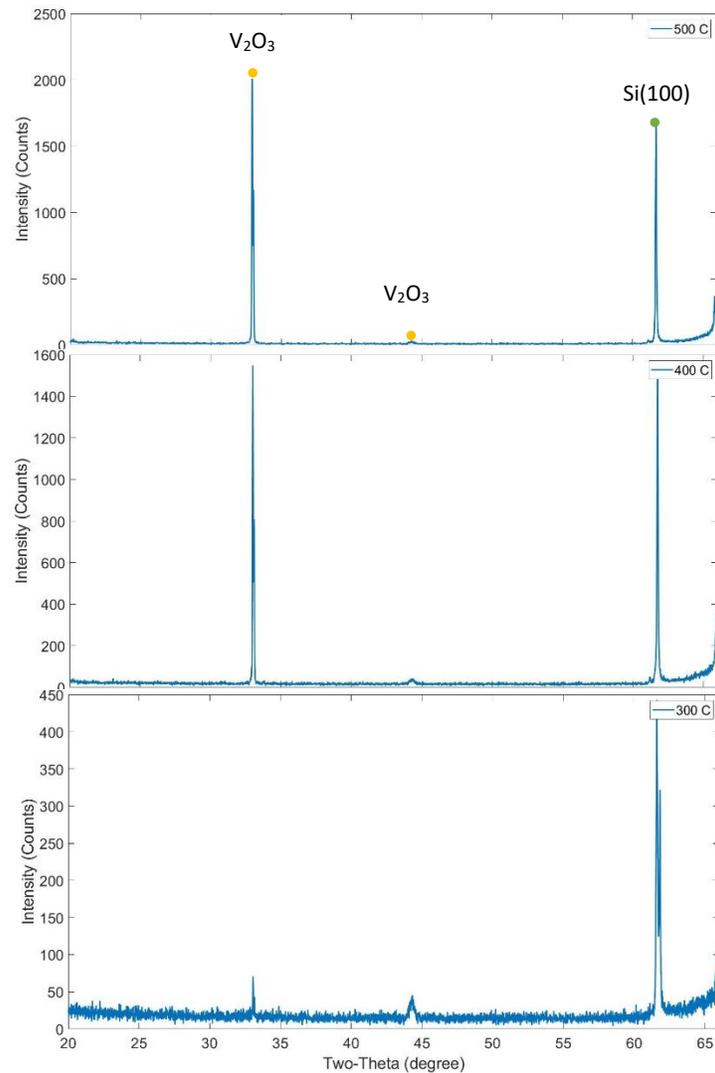

Figure 11. XRD result for sputtered film with Rfo 0.058 at temperature from 300°C (bottom), 400°C (middle) and 500°C (top)

## 5. CONCLUSION

In this article, the crystallographic and electrical properties of $VO_2$ were reviewed to show the metal-insulator phase transition of the film at temperature 68 °C. The proposed $VO_2$ based metamaterial structure was simulated with HFSS to demonstrate the emittance tunability ($\Delta\varepsilon$) of rutile and monoclinic states of $VO_2$. We summarized the recent progress in the in the deposition of thermochromic $VO_2$ on glass and silicon substrate by focusing on changing the partial pressure of the oxygen and argon. Furthermore, we sputtered VOx by $V_2O_5$ target with Rfo between 0.042 to 0.6 at temperature 300°C and higher and measured it with XRD method, which resulted in forming the $V_2O_3$ on the thick oxide on silicon wafer. For future work, it is recommended to sputter the thin film VOx on sapphire or silicon substrate with $V_2O_3$, since $VO_2$ grows smoothly on sapphire in comparison with the thick thermal $SiO_2$. Thin films thermochromic vanadium dioxide is currently attracting much attention in areas such as microelectronics and spacecraft control systems for their electronic behaviors in MIT.

## 6. ACKNOWLEDGEMENT

This work was performed in part at the Cornell NanoScale Facility, a member of the National Nanotechnology Coordinated Infrastructure (NNCI), which is supported by the National Science Foundation (Grant NNCI-2025233). Also, we appreciate the Cornell Center for Materials Research Shared Facilities which are supported through the NSF MRSEC program (DMR-1719875). We would like to thank the NSF Industry/University Cooperative Research Center for Metamaterials to support this project.

[33] Hassein-Beya A. L. S., Lafanea S., Tahib H., Hassein-Bey A., Djafara A. A., Abdelli-Messacia S. and Benamard M. E. A., "Substrate Effect on Structural, Microstructural and Elemental Microcomposition of Vanadium Dioxide Thin Film," *Modern Arabic Reviewof Fundamental & Applied Physics,* vol. 1, no. 2, pp. 24 -28, (2017).

[34] Tangirala M., Zhang K., Nminibapiel D., Pallem V., Dussarrat C., Cao W., Adam T. N., Johnson C. S., Elsayed-Ali H. E. and Baumgart H., "Physical Analysis of VO2 Films Grown by Atomic Layer Deposition and RF Magnetron Sputtering," *ECS Journal of Solid State Science and Technology,* vol. 3, no. 6, pp. N89-N94, (2014).

[35] Wang X., Guo Z., Gao Y. and Wang J., "Atomic layer deposition of vanadium oxide thin films from tetrakis(dimethylamino)vanadium precursor," *JMR EARLY CAREER SCHOLARS IN MATERIALS SCIENCE ANNUAL ISSUE,* vol. 32, no. 1, (2017).

[36] Currie M., Mastro M. A. and Wheeler V. D., "Atomic Layer Deposition of Vanadium Dioxide and a Temperature-dependent Optical Model," *Journal of Visualized Experiments,* vol. 135, no. e57103, (2018).

[37] Peter A. P., Toeller M., Adelmann C. and Schaekers M., "Process study and characterization of VO2 thin films synthesized by ALD using TEMAV and O3 precursors," *ECS Journal of Solid State Science and Technology,* vol. 1, no. 4, pp. 169-174, (2012).

[38] Mirbagheri G., "Hyperbolic Metamaterial Filter for Angle-Independent TM Transmission in the Infrared Regime," *NanoMeter The Newsletter of the Cornell Nano Scale Facility,* vol. 29, no. 2, p. 17, (2020).